\documentclass[aps,pra,amsmath,amssymb,preprintnumbers,showpacs,twocolumn]{revtex4-1} 

\usepackage[latin1]{inputenc}
 \usepackage{bm}
\usepackage{multirow,amssymb,amsbsy,amsmath}
\usepackage{amsthm}
\usepackage[pdftex]{graphicx}
 \usepackage{pgfplots}

\usepackage{verbatim}
\usepackage{color}
\makeatletter
\usepackage{pifont}

\usepackage{epstopdf}

\begin{document}

\title{Determining the squeezing in multimode fields using nonlocal quantum probes}

\author{Steffen Wi\ss mann}
\author{Heinz-Peter Breuer}
\affiliation{Physikalisches Institut, Universit\"{a}t Freiburg, Hermann-Herder-Stra\ss e 3, D-79104 Freiburg, Germany}

\begin{abstract}
We present a scheme allowing to access the squeezing parameter of multimode 
fields by means of the dynamics of nonlocal quantum probes. The model under 
consideration is composed of two two-level systems which are coupled locally to 
an environment consisting of nonlocally correlated field modes given by two-mode 
Gaussian states. Introducing independently switchable interactions, one observes 
revivals of nonlocal coherences of the two-qubit system which are unambiguously 
connected to the squeezing parameter of the Gaussian environmental states. 
Thus, the locally interacting two two-level systems represent a dynamical quantum 
probe for the squeezing in multimode fields. It is finally demonstrated that perfectly 
reviving nonlocal coherences also persists for nonentangled correlated field 
modes and an explanation for this phenomenon is presented by connecting it to 
the correlation coefficient of the environmental coupling operators.
\end{abstract}


\maketitle

\section{Introduction}\label{sec:intro}

In almost any experiment and application of interacting many-body quantum systems only a small number of degrees of freedom can be controlled and manipulated while the rest remains experimentally inaccessible. The theory of open quantum systems \cite{BreuerBuch} provides a different view on this fact: The controllable quantities define an open system whose dynamics is substantially affected by the inaccessible degrees of freedom constituting the environment. The environmental properties are thus imprinted on the dynamics of the open system to a certain degree making the open system a quantum probe for the environment.

Recently, the use of quantum probes has been demonstrated experimentally ranging from all-optical setups \cite{Nonlocal,PhotonCorrAndrea} to quantum many-body systems \cite{Realization}. In addition, several theoretical studies have shown quantum probing strategies in various other quantum systems \cite{BECPinja0,BECPinja1,BECPinja2,NonlocalTheo,IonCrystals,IonCrystals2,IonCrystalsMassimoPRA}. The extraction of information about the complex environment employing a quantum probe is frequently connected to non-Markovian dynamics using the concept of the information flux between the open system and its environment \cite{Measure,MeasurePaper}. More precisely, the environmental properties are revealed by the non-Markovianity of the open system dynamics \cite{BECPinja0,NonlocalTheo,IonCrystalsMassimoPRA}.

Several experiments \cite{Nature,SingleQubitExp,LocRep} have been performed quantifying non-Markovian behavior in terms of the flow of information between the open system and its environment. Apart from long-lasting, non-negligible system-environment correlations inducing memory effects, it has been shown theoretically as well as experimentally that they can also be caused by nonlocal environmental
correlations \cite{NonlocalTheo,Nonlocal}. The first model studied in Ref.\,\cite{NonlocalTheo} illustrates this effect by means of an open system consisting of two qubits coupled locally to a multimode field in a nonlocally correlated
initial state. Here, we consider this model with independently switchable interactions, which is inspired by the idea of having controllable quantum sensors probing the system of interest, and provide the correct expressions for the quantum dynamical map for environmental states given by a product of two-mode Gaussian states. For suitably adjusted parameters and consecutively applied interactions, there are almost perfectly reviving nonlocal coherences identifying nonlocal memory effects. 

We then demonstrate how these nonlocal memory effects are influenced by the properties of the environmental multimode field imprinting unique signatures on the dynamics of the open system. It can be shown that the strength of the renascent coherences, corresponding to the recovery of entanglement of maximally entangled two-qubit Bell states, is unambiguously connected to the squeezing in two-mode Gaussian states as a function of the interaction length of the subsequently applied local interactions. By monitoring the rephasing of the nonlocal coherences for various time durations of the interaction one is able to reveal the squeezing of the multimode field. As the rephasing is inherently linked to a backflow of information the dynamics of the open system thus provides a non-Markovian dynamical quantum probe for these environmental properties. 

We finally demonstrate that entanglement is not necessary for the occurrence of perfectly reviving nonlocal coherences by considering nonentangled, mixed two-mode Gaussian states showing even stronger memory effects. By means of an approximated dynamics, neglecting the free evolution, an explanation for the phenomenon of nonlocal memory effects along with a necessary and sufficient condition for the properties of the two-mode Gaussian states for sufficiently short interactions is provided by linking it to the magnitude of the correlation coefficient of the environmental coupling operators. 

\section{Physical model}\label{sec:model}
\begin{figure}[b]
    \centering
     \includegraphics[width=0.3\textwidth]{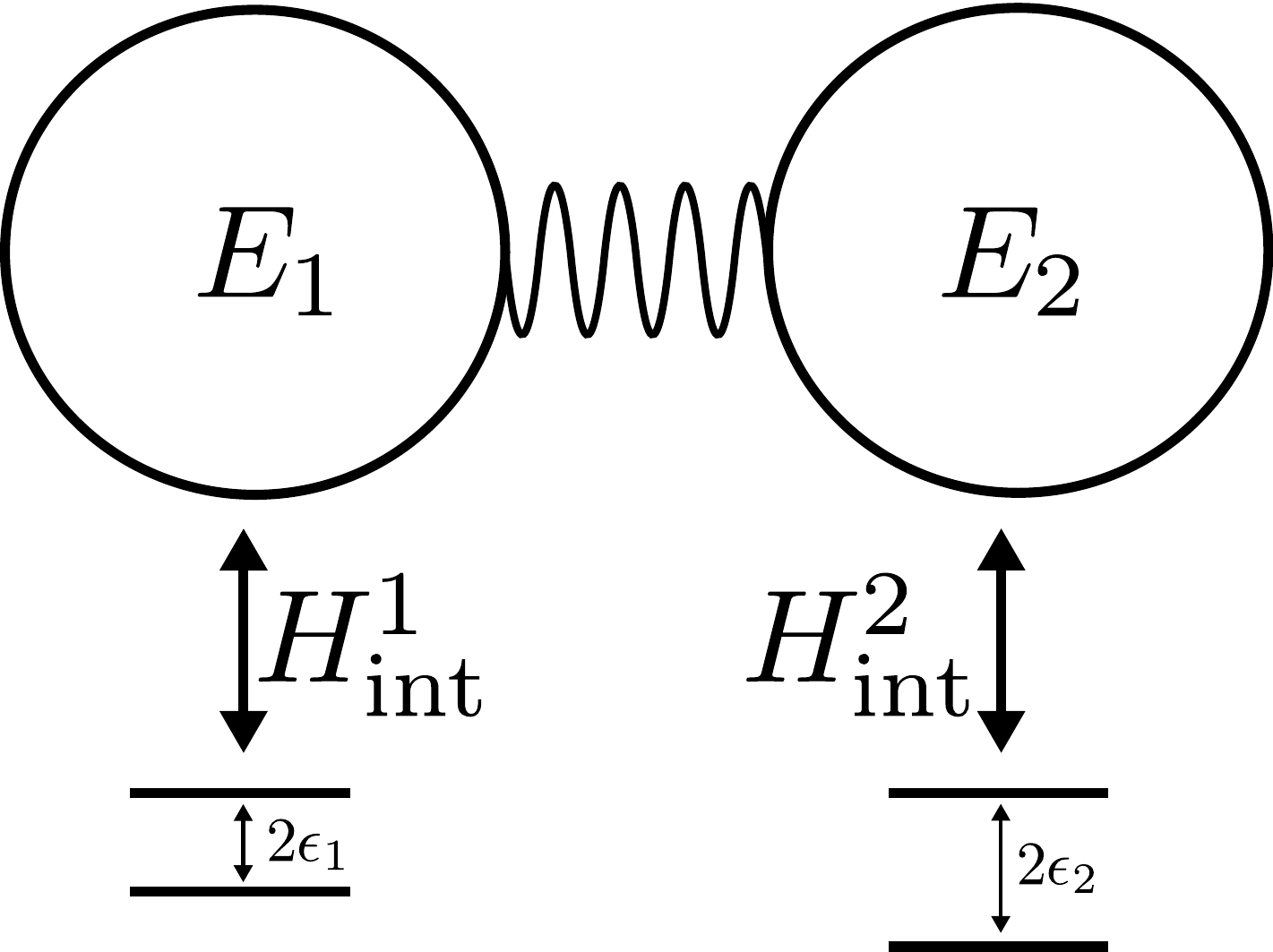}\\
    \caption{Schematic picture of two two-level systems coupled locally to a correlated multimode environment. The environment is initially in a product state of two-mode Gaussian states correlating pairs of modes of the two baths. The local interactions can be turned on and off independently.} \label{fig:Schema}
 \end{figure}

We investigate the model of two two-level systems coupled to a bosonic environment which was studied previously regarding the occurrence of nonlocal memory 
effects \cite{NonlocalTheo}. Each of the two qubits interacts locally with its own multimode bosonic environment, where the two environments are initially in a correlated total state (see Fig.\,\ref{fig:Schema}). In fact, one chooses for the initial state a product of two-mode Gaussian states correlating pairs of modes of the two bosonic environments.

The Hamiltonian of the total system is given by
\begin{align}
H&=H_0+H_{int}(t)~, \qquad H_0=\sum_{i=1}^2(H_S^i+H_E^i)~,
\end{align}
with
\begin{align}
H_S^i&=\epsilon_i\hat{\sigma}_z^i~,\qquad H_E^i=\sum_k\omega_k^i {\hat{b}_k}^{i\dag} \hat{b}_k^i~.
\end{align}
 Here, ${\hat{b}_k}^{i(\dag)}$ refer to the annihilation (creation) operator of the $k$th mode of bath $i=1,2$ which satisfies the familiar commutation relation $[\hat{b}_k^i,{\hat{b}_l}^{j\dag}]=\delta_{kl}\delta_{ij}$. Furthermore,  $\hat{\sigma}_z^i$ denotes the usual Pauli matrix and the energy gap $2\epsilon_i$ of the two qubits can be arbitrary. The interaction Hamiltonian is composed of local interactions, i.e. $H_{int}(t)=\sum_{i}H_{int}^i(t)$ which obey
\begin{equation}\label{eq:Hint}
 H_{int}^i(t)=\chi_i(t)\sum_k \hat{\sigma}_z^i\otimes(g_k^i {\hat{b}_k}^{i\dag} + {g_k^i}^* \hat{b}_k^i)~.
\end{equation}
The coefficient $g_k^i$ denotes the strength of the coupling between subsystem $i$ and its bath mode $k$. Without loss of generality we assume that the coupling strengths are real-valued, i.e. $g_k^i\in\mathbb{R}$ for $i=1,2$ and all $k$. The function $\chi_i(t)$ is given by
\begin{equation}
\chi_i(t)=\begin{cases}
                                          1~ ,~ t\in[t_i^s,t_i^f]\\ 0~ ,~ \text{else}
                                         \end{cases}
 \end{equation}
for some $t_i^f>t_i^s>0$. It models the turning-on and -off of the local interactions of subsystem $i$ at time $t_i^s$ and $t_i^f$, respectively. The duration of the local interactions and the turning-on can be varied independently for both subsystems so that it is possible to tune continuously from simultaneous to a successive application of the interactions. This setup is motivated by the idea of having controllable quantum sensors which can be brought in contact with a complex environment for a certain time interval, thereby acting as a quantum probe for environmental properties. Without loss of generality we assume $t_1^s\leq t_2^s$.

The essential part of the Hamiltonian is a well-known, simple example of the spin-boson model leading to decoherence  of the two-level system \cite{DiVincenzo}. The  dynamics of the model is conveniently solved in the interaction picture. Turning to this picture the interaction Hamiltonian transforms into
\begin{equation}
H_{int}^I(t)=\sum_{j}\chi_j(t)\hat{\sigma}_z^j \otimes \sum_k (g_k^j e^{i\omega_k^j t}{\hat{b}_k}^{j\dag} + {g_k^j}^* e^{-i\omega_k^j t} \hat{b}_k^j).
\end{equation}
Due to the bosonic commutation relation the commutator of the Hamiltonian at different times is a c-number function, i.e.
\begin{align}
 &[H_{int}^I(t),H_{int}^I(t')]=-2i \phi(t-t')~,
\end{align}
where $\phi(t-t')=\sum_{j,k}\chi_j(t)\chi_j(t') |g_k^j|^2 \sin[\omega_k^j(t-t')]$. It is well known \cite{BreuerBuch} that the time evolution operator in the interaction picture is then given by
\begin{align}
 U_I(t)&=\mathrm{T} \exp\left[-i\int_0^t \mathrm{ds} H_{int}^I(s)\right] \nonumber\\
 &=\exp\left[i\int_0^t \mathrm{ds}\int_0^t \mathrm{ds'}\phi(s-s')\Theta(s-s')\right]\nonumber\\
&\qquad\cdot \exp\left[-i\int_0^t \mathrm{ds} H_{int}^I(s)\right],
\end{align}
where $\mathrm{T}$ denotes the chronological time-ordering.
The evolution operator thus consists of an overall phase factor $d(t)\equiv\exp[i\int_0^t \mathrm{ds}\int_0^t \mathrm{ds'}\phi(s-s')\Theta(s-s')]$ and a nontrivial unitary evolution operator $V(t)\equiv\exp[-i\int_0^t \mathrm{ds} H_{int}^I(s)]$ which can be rewritten as
\begin{align}
V(t)&=\exp\left[\sum_{j,k}\hat{\sigma}_z^j \otimes (\beta_k^j(t){\hat{b}_k}^{j\dag} - \beta_k^j(t)^*\hat{b}_k^j)\right]~,
\end{align}
where
\begin{equation}
 \beta_k^j(t)=\frac{g_k^j}{\omega_k^j}\,e^{i\omega_k^jt_j^s}\Bigl(1-e^{i\omega_k^j t_j(t)}\Bigr)~,
\end{equation}
and $t_j(t)\equiv\int_0^t \mathrm{ds} \chi_j(s)$. Here, we used that $\int_0^t\mathrm{ds}\chi_j(s)e^{i\omega_k^j s}=i e^{i\omega_k^jt_j^s}(1-e^{i\omega_k^j t_j(t)})/\omega_k^j$. Note that the phase factor $e^{i\omega_k^jt_j^s}$, taking into account the influence of the free evolution prior the turning-on of the interaction, is missing in Ref.\,\cite{NonlocalTheo}. One concludes that the operator $V(t)$ is a product of two-mode displacement or Weyl operators. The local unitaries $V_j(t)$ ($V(t)\equiv V_1(t)V_2(t)$) act therefore according to
\begin{align}
V_j(t)|0\rangle\otimes|\eta\rangle&=|0\rangle\otimes\prod_k D(-\beta_k^j(t))|\eta\rangle~, \\
V_j(t)|1\rangle\otimes|\eta\rangle&=|1\rangle\otimes\prod_k D(+\beta_k^j(t))|\eta\rangle~,
\end{align}
where $D$ denotes the (single-mode) displacement operator and $|0\rangle$, $|1\rangle$ refer to the ground and excited state of the two-level system, respectively. Moreover, $|\eta\rangle$ is an arbitrary pure state of the environment. Finally, the time evolution operator in the Schr\"odinger picture is given by
\begin{equation}
 U(t)=e^{-iH_0 t}U_I(t)~.
\end{equation}
Thus, the time evolution of the initially factorizing state
\begin{align}
 |\psi(0)\rangle&=|\psi_{12}\rangle\otimes|\eta_{12}\rangle~, \\
\end{align}
where $|\psi_{12}\rangle$ refers to any pure state of the two-qubit system, i.e.
\begin{align}
|\psi_{12}\rangle&=a_{00}|00\rangle+a_{01}|01\rangle+a_{10}|10\rangle+a_{11}|11\rangle~, \\
\intertext{and}
|\eta_{12}\rangle&=\bigotimes_k|\eta_{12}^k\rangle~,\label{eq:eta}
\end{align}
is a product state of arbitrary two-mode states $|\eta_{12}^k\rangle$ of the $k$th mode of bath $1$ and $2$, reads
\begin{equation}
 |\psi(t)\rangle=e^{-iH_0 t}d(t)\sum_{m,n=0}^1 a_{mn}|mn\rangle\otimes|\eta_{12}^{mn}(t)\rangle~.
\end{equation}
Here, the time-evolved environmental states are given by 
\begin{align}
\lefteqn{|\eta_{12}^{mn}(t)\rangle}\\
&\equiv\bigotimes_k D\bigl((-1)^{m+1}\beta_k^1(t)\bigr)\otimes D\bigl((-1)^{n+1}\beta_k^2(t)\bigr)|\eta_{12}\rangle~.\nonumber
\end{align}
The reduced state of the two two-level systems is then obtained by taking the partial trace over the environmental degrees of freedom which yields
\begin{align}\label{eq:genstate}
&\rho_S^{12}(t)\\
=&\sum_{m,n,r,s=0}^1 e^{it\bigl([(-1)^m-(-1)^{r}]\epsilon_1+[(-1)^n-(-1)^{s}]\epsilon_2\bigr)}\nonumber\\
&\qquad\qquad\qquad a_{mn}a_{rs}^* \cdot\langle \eta_{12}^{rs}(t)|\eta_{12}^{mn}(t)\rangle \cdot|mn\rangle\langle rs|\nonumber\\\nonumber\\
=&\left(\begin{matrix}|a_{11}|^2 & a_{11}a_{10}^*\tilde{\kappa}_2(t) & a_{11}a_{01}^*\tilde{\kappa}_1(t) & a_{11}a_{00}^*\kappa_{12}(t) \\
 & |a_{10}|^2 & a_{10}a_{01}^*\Lambda_{12}(t) & a_{10}a_{00}^*\kappa_1(t) \\
 & &|a_{01}|^2 & a_{01}a_{00}^*\kappa_2(t)\\
 \text{c.c.}& & &|a_{00}|^2 \\ \end{matrix}\right)~,\nonumber
\end{align}
where
\begin{align}\label{eq:coh}
 \kappa_1(t)&=e^{-2i\epsilon_1t}\langle\eta_{12}^{00}(t)|\eta_{12}^{10}(t)\rangle~, \\
 \kappa_2(t)&=e^{-2i\epsilon_2t}\langle\eta_{12}^{00}(t)|\eta_{12}^{01}(t)\rangle~, \\
\tilde{\kappa}_1(t)&=e^{-2i\epsilon_1t}\langle\eta_{12}^{01}(t)|\eta_{12}^{11}(t)\rangle~, \\
\tilde{\kappa}_2(t)&=e^{-2i\epsilon_2t}\langle\eta_{12}^{10}(t)|\eta_{12}^{11}(t)\rangle~, \\
 \kappa_{12}(t)&=e^{-2i(\epsilon_1+\epsilon_2)t}\langle\eta_{12}^{00}(t)|\eta_{12}^{11}(t)\rangle~, \\
 \Lambda_{12}(t)&=e^{-2i(\epsilon_1-\epsilon_2)t}\langle\eta_{12}^{01}(t)|\eta_{12}^{10}(t)\rangle~,\label{eq:cohl}
\end{align}
and 
\begin{align}
 &\langle\eta_{12}^{rs}(t)|\eta_{12}^{mn}(t)\rangle\nonumber\\
&=\prod_k\langle\eta_{12}^k|\left[D\bigl((-1)^{r+1}\beta_k^1(t)\bigr)\otimes D\bigl((-1)^{s+1}\beta_k^2(t)\bigr)\right]^\dag\nonumber\\
&\hspace{1.5cm}\left[D\bigl((-1)^{m+1}\beta_k^1(t)\bigr)\otimes D\bigl((-1)^{s+1}\beta_k^2(t)\bigr)\right]|\eta_{12}^k\rangle\nonumber\\
&=\prod_k \chi_k^{rsmn}~.
\end{align}
Using the identities $D^\dag(\alpha)=D(-\alpha)$ and $D(\alpha)D(\beta)=e^{i\textrm{Im}(\alpha\beta^*)}D(\alpha+\beta)$ for displacement operators one obtains
\begin{align}\label{eq:twomode}
\lefteqn{\chi_k^{rsmn}}\\
&=\mathrm{tr}\Bigl\{\exp\left[\sum_{j=1}^2\gamma_{k,rsmn}^j(t) {b_k^j}^\dag-\gamma_{k,rsmn}^j(t)^*b_k^j\right]\rho_{E,k}\Bigr\}~,\nonumber
\end{align}
with $\rho_{E,k}=|\eta_{12}^k\rangle\langle\eta_{12}^k|$ and 
\begin{align}
\gamma_{k,rsmn}^1(t)&\equiv\{(-1)^r-(-1)^m\}\beta_k^1(t)~,\\
\gamma_{k,rsmn}^2(t)&\equiv\{(-1)^s-(-1)^n\}\beta_k^2(t)~.
\end{align}
Hence, $\chi_k^{rsmn}$ is the Wigner characteristic function of the pure state $|\eta_{12}^k\rangle$ which is easily determined for two-mode Gaussian states. The previous derivation assumed for the sake of clarity pure environmental states which can however easily be generalized to arbitrary states $\rho_{E,k}$ of the $k$th bath modes.

\section{Coherence factors for two-mode Gaussian states}\label{sec:2modeGauss}
In the following we present the explicit expressions for the coherence factors \eqref{eq:coh}$-$\eqref{eq:cohl} if the environmental state $\rho_{E,k}$ of the $k$th bath modes is chosen to be a two-mode Gaussian state whose covariance matrix is in standard form. These states are distinguished among other continuous variable system states due to their experimental relevance and simple mathematical structure. More precisely, they arise for example as states of the light field of lasers and are completely determined by the first and second moments of canonical operators. Moreover, extending the concept of quantum discord to continuous variable systems \cite{CVQuantumDiscord,CVQuantumDiscord2}, it has been shown that all two-mode Gaussian states which do not factorize have nonzero quantum discord. Without loss of generality one may assume that the Gaussian state has zero mean as this can always be achieved applying local unitary operations \cite{GaussianIllumi,GaussianIndia}. This does, however, not change the quantum correlations of the two-mode state measured by the quantum discord we are mainly interested in. 
 
We recall that a state of a continuous variable system 
$\rho\in\mathcal{S}(L^2(\mathbb{R}^n))$ is an $n$-mode Gaussian state if and only if for all $\vec{x},\vec{y}\in\mathbb{R}^n$ the observable $\hat{Y}\equiv\sum_{j=1}^n(x_j\hat{p}_j-y_j\hat{q}_j)$ has a normal distribution on $\mathbb{R}$ in the state $\rho$ \cite{GaussianIndia}, where 
\begin{equation}
 \hat{q}_j=\frac{1}{\sqrt{2}}(\hat{b}_j+\hat{b}_j^\dag)~,\qquad\hat{p}_j=\frac{-i}{\sqrt{2}}(\hat{b}_j-\hat{b}_j^\dag)
\end{equation}
define the canonical position and momentum operators. That is, one finds for the characteristic function
\begin{align}\label{eq:chiGen}
 \chi_{\rho}^t(\vec{z})&\equiv\mathrm{Tr}(\rho \exp[-i t \hat{Y}])\nonumber\\
 &=\exp\left[-it(\boldsymbol{l}^T\vec{x}-\boldsymbol{m}^T\vec{y})-\frac{t^2}{2}\vec{w}^T\boldsymbol{S}\vec{w}\right]~,
\end{align}
where $z_j=x_j+i y_j$ for all $j$ and
$\vec{w}^T=(y_1,x_1,\dots,y_n,x_n)$ and $\boldsymbol{l}_i=\langle\hat{p}_i\rangle$, $\boldsymbol{m}_i=\langle\hat{q}_i\rangle$ denote the mean position and momentum. Moreover, the $(2n\times 2n)$-matrix $\boldsymbol{S}$ is the covariance matrix of the operator $\hat{X}'=(\hat{q}_1,-\hat{p}_1,\dots,\hat{q}_n,-\hat{p}_n)$, i.e.
\begin{align}
\boldsymbol{S}&=\boldsymbol{\sigma}_{X'}\equiv \left(\Bigl(\tfrac{1}{2}\langle\{\hat{X}'_i,\hat{X}'_j\}\rangle-\langle\hat{X}'_i\rangle\langle\hat{X}'_j\rangle\Bigr)_{ij}\right)~.\label{eq:covS}
\end{align}
Equivalently, one may replace $\boldsymbol{S}$ by $\Omega_n\boldsymbol{\sigma}_X\Omega_n^T$ \cite{Olivares,ReviewGaussian}, where the covariance matrix is given with respect to the usual canonical operators $\hat{X}=(\hat{q}_1,\hat{p}_1,\dots,\hat{q}_n,\hat{p}_n)$ and 
\begin{equation}\label{eq:symplForm}
 \Omega_n=\oplus_{k=1}^n \omega\qquad \text{with} \qquad \omega=\left(\begin{matrix}0&1\\-1&0\end{matrix}\right)
\end{equation}
defines the symplectic form encoding the canonical commutation relations.

For $t=\sqrt{2}$ the operator $\exp[-i t \hat{Y}]$ is called the Weyl operator $\mathcal{W}(\vec{z})$
\begin{align}\label{eq:Weyl}
 \mathcal{W}(\vec{z})&=\exp\left[\sum_{j=1}^n(z_j\hat{b}_j^\dag-z_j^*\hat{b}_j)\right].
\end{align}
One can show \cite{GaussianIndia,SimonGaussian} that the right hand side of Eq.\,\eqref{eq:chiGen} defines the characteristic function of an $n$-mode Gaussian state for some $\boldsymbol{l},\boldsymbol{m}\in\mathbb{R}^n$ and $\boldsymbol{S}\geq0$ if and only if
\begin{align}\label{eq:conCova}
\boldsymbol{S}+\frac{i}{2}\Omega_n\geq0~.
\end{align}
This condition is sometimes called the Robertson-Schr\"odinger uncertainty relation and is a direct consequence of the Schr\"odinger uncertainty relation and Williamson's theorem \cite{Williamson} which states that any real-valued, symmetric and positive matrix can be transformed into diagonal form by an appropriate symplectic operation \cite{SimonGaussian}. Note that Eq.\,\eqref{eq:conCova} implies positivity of $\boldsymbol{S}$ and that one has $\boldsymbol{S}+(i/2)\Omega_n\geq0$ if and only if $\boldsymbol{S}-(i/2)\Omega_n\geq0$. 

Now, suppose that the environmental states $\rho_{E,k}$ for all $k$ are two-mode Gaussian states with zero mean. According to Eq.\,\eqref{eq:chiGen}, Eq.\,\eqref{eq:twomode} is then given by
\begin{align}\label{eq:char1}
 &\chi_k^{rsmn}\bigl((\gamma_{k,rsmn}^1(t),\gamma_{k,rsmn}^2(t))\bigr)\nonumber\\
 &=\exp\left[-\vec{\lambda}_{k,rsmn}(t)^T\boldsymbol{S}_k\vec{\lambda}_{k,rsmn}(t)\right]~,
\end{align}
where 
\begin{align}
\vec{\lambda}_{k,rsmn}(t)\equiv\left(\begin{matrix}\textrm{Im}(\gamma_{k,rsmn}^1(t))\\\textrm{Re}(\gamma_{k,rsmn}^1(t))\\\textrm{Im}(\gamma_{k,rsmn}^2(t))\\\textrm{Re}(\gamma_{k,rsmn}^2(t))\end{matrix}\right)~,
\end{align}
and $\boldsymbol{S}_k$ satisfies Eq.\,\eqref{eq:conCova}. We point out that the covariance matrix which is considered in Ref.\,\cite{NonlocalTheo} violates Eq.\,\eqref{eq:conCova} for any $c\neq0$\,. 

For any two-mode covariance matrix $\boldsymbol{S}$ there exist local symplectic operations such that the expectation values $\langle\{\hat{q}_i,\hat{p}_j\}\rangle$ are removed \cite{GaussianZoller,GaussianSimon}. This implies that $\boldsymbol{S}$ is transformed into the so called standard form
\begin{align}
\boldsymbol{S}_{\mathrm{sf}}&\equiv\left(\begin{matrix} a&0&c_+&0\\0&a&0&c_-\\c_+&0&b&0\\0&c_-&0&b                                                                                                                                                                                                                                                                                                                       \end{matrix}\right)~,\label{eq:standardform1}
\end{align}
where $a,b,c_\pm\in\mathbb{R}$ and $a,b\geq1/2$. We remark that the local symplectic operations correspond to special local unitaries on the level of density operators \cite{GaussianIndia,CVQuantumDiscord2} so that the quantum discord of two-mode states is invariant under such a transformation of the covariance matrix. As a covariance matrix in standard form has vanishing cross-covariances $\langle\{\hat{q}_i,\hat{p}_j\}\rangle$, one directly sees from the definition that such a matrix, associated to the set of operators $\hat{X}'=(\hat{q}_1,-\hat{p}_1,\hat{q}_2,-\hat{p}_2)$, equals the covariance matrix $\boldsymbol{\sigma}_X$ obtained for $\hat{X}=(\hat{q}_1,\hat{p}_1,\hat{q}_2,\hat{p}_2)$.

The coherence factors defined in Eqs.\,\eqref{eq:coh}-\eqref{eq:cohl} can now be written as products of characteristic functions, i.e. 
\begin{align}
 \kappa_1(t)&=e^{-2i\epsilon_1t}\nonumber\\
 &\qquad\cdot\exp\left[-\sum_k \vec{\lambda}_{k,0010}(t)^T \boldsymbol{S}_{k} \vec{\lambda}_{k,0010}(t)\right]~,\label{eq:kappa1P} \\
 \kappa_2(t)&=e^{-2i\epsilon_2t}\nonumber\\
 &\qquad\cdot \exp\left[-\sum_k \vec{\lambda}_{k,0001}(t)^T \boldsymbol{S}_{k} \vec{\lambda}_{k,0001}(t)\right]~,\label{eq:kappa2P} \\
\kappa_{12}(t)&=e^{-2i(\epsilon_1+\epsilon_2)t}\nonumber\\
&\qquad\cdot \exp\left[-\sum_k \vec{\lambda}_{k,0011}(t)^T \boldsymbol{S}_{k} \vec{\lambda}_{k,0011}(t)\right]~,\label{eq:kappa12P} \\
 \Lambda_{12}(t)&=e^{-2i(\epsilon_1-\epsilon_2)t}\nonumber\\
 &\qquad \cdot \exp\left[-\sum_k\vec{\lambda}_{k,0110}(t)^T \boldsymbol{S}_{k} \vec{\lambda}_{k,0110}(t)\right]~,\label{eq:lambda12P} 
\end{align}
where $\tilde{\kappa}_j(t)=\kappa_j(t)$ and $\boldsymbol{S}_k$ refers to the covariance matrix of a zero-mean two-mode Gaussian state. Henceforth, we assume that the Gaussian states are identical for all modes, i.e. $\boldsymbol{S}_k=\boldsymbol{S}$ for all $k$ and are in standard form. After performing the continuum limit for an ohmic spectral density $J_j(\omega)=\alpha_j \omega \exp[-\omega/\omega_c]$ with equal cutoff frequency $\omega_c$ but different couplings $\alpha_j$ for the two bosonic baths, the exponential functions in Eqs.\,\eqref{eq:kappa1P}-\eqref{eq:lambda12P} can be evaluated for a general covariance matrix in standard form employing Laplace transforms. One obtains expressions containing the Laplace transform $\mathcal{L}$ of $(1-\cos(y t))/t$ and $\sin$-modulated functions in the exponent which can be evaluated using standard techniques. More precisely, by means of
\begin{align}
&\mathcal{L}\left\{(1-\cos(y t))/t\right\}(s)=\frac{1}{2}\log\left[1+\frac{y^2}{s^2}\right]~, \\
&\mathcal{L}\left\{\sin(x t)\tfrac{\sin(y t)}{t}\right\}(s)=\frac{1}{4}\log\left[\frac{(y+x)^2+s^2}{(y-x)^2+s^2}\right]~,
\end{align}
for $s\neq0$ one obtains for the coherence factors \eqref{eq:kappa1P}-\eqref{eq:lambda12P}
\begin{align}
  \kappa_1(t)&=e^{-2i\epsilon_1t}\left(1+\omega_c^2t_1(t)^2\right)^{-4 a \alpha_1}~,\label{eq:kappa1}\\
  \kappa_2(t)&=e^{-2i\epsilon_2t}\left(1+\omega_c^2t_2(t)^2\right)^{-4 b \alpha_2}~,\label{eq:kappa2}\\
  \kappa_{12}(t)&=\kappa_1(t)\kappa_2(t)f(t)~,\label{eq:kappa12}\\
\Lambda_{12}(t)&=\kappa_1(t)\kappa_2^*(t) / f(t)~,\label{eq:lambda12}
\end{align}
where
\begin{widetext}
\begin{align}
f(t)&=\left(\frac{(1+\omega_c^2{t_2^s}^2)(1+\omega_c^2(t_1(t)-t_2(t)-t_2^s)^2)}{(1+\omega_c^2(t_1(t)-t_2^s)^2)(1+\omega_c^2(t_2(t)+t_2^s)^2)}\right)^{4c_-\sqrt{\alpha_1\alpha_2}}\nonumber\\
&\qquad\cdot\left(\frac{\bigl(1+\omega_c^{2}(t_1(t)-t_2^s)^2\bigr)\bigl(1+\omega_c^{2}(t_2(t)+t_2^s+t_1(t))^2\bigr)}{\bigl(1+\omega_c^{2}(t_1(t)+t_2^s)^2\bigr)\bigl(1+\omega_c^{2}(t_2(t)+t_2^s-t_1(t))^2\bigr)}\right)^{2(c_--c_+)\sqrt{\alpha_1\alpha_2}}\label{eq:funcf}
\end{align}
 \end{widetext}
given a covariance matrix in standard form \eqref{eq:standardform1} with real-valued coefficients $a$, $b$ and $c_\pm$. Here, we have set $t_1^s=0$ for simplicity. The time $t_2^s$, at which the interaction of the second spin with its bath is turned on, remains however arbitrary. 

\section{Squeezed thermal states}\label{sec:two-modeStates}
Having derived the expressions for the coherence factors assuming two-mode Gaussian states in standard form  \eqref{eq:standardform1} with zero means, we now turn to a relevant subclass of these states for which we have $c_+=-c_-$. We consider squeezed thermal states (STS) which are obtained from thermal states, described by a diagonal covariance matrix, by applying only linear and bilinear interactions of modes. More precisely, they are generated by the (two-mode) squeezing operator $S_2(r)=\exp[r (\hat{b}_1^\dag \hat{b}_2^\dag- \hat{b}_1\hat{b}_2)]$ with $r\in\mathbb{R}$ acting on the tensor product of thermal states with mean photon number $N_1$ and $N_2$ for the modes associated to the annihilation (creation) operators $\hat{b}_1^{(\dag)}$ and $\hat{b}_2^{(\dag)}$, respectively \cite{SqueezedStates}. The variable $r$ is referred to as the squeezing parameter.  Values of about $r=5$ can be realized in experiments \cite{Werner}. The associated covariance matrix $\boldsymbol{\sigma}_{X,r}^\mathrm{STS}$ is in standard form and determined by
\begin{align}\label{eq:squeezedThermala}
a&=\frac{1}{2}\cosh(2r)+N_1\cosh^2(r)+N_2\sinh^2(r)~,\\
b&=\frac{1}{2}\cosh(2r)+N_2\cosh^2(r)+N_1\sinh^2(r)~,\label{eq:squeezedThermalb}\\
c_+&=\frac{1}{2}(1+N_1+N_2)\sinh(2r)=-c_-~.\label{eq:squeezedThermalc}
\end{align}

As one can see, the squeezing operator is an active device meaning that it adds energy to the initial states. The difference in the mean photon number is, however, a constant of motion so that the squeezing operator adds or removes excitations to the same degree to either modes. The assumption of equal covariance matrices $\boldsymbol{S}_k$ for all pairs of modes $k$ amounts to a mode-dependent temperature such that the ratio $(\hbar \omega)/(k_B T)$ is constant. For high temperatures and a small cutoff frequency this is approximately satisfied.

The entanglement properties of the squeezed thermal states are uncovered using the PPT-criterion which has been proven to be a necessary and sufficient entanglement criterion for two-mode Gaussian states \cite{GaussianZoller,GaussianSimon}. Applying this criterion one finds that a squeezed thermal state is separable if and only if the following inequality is satisfied:
\begin{equation}\label{eq:PPT}
 \cosh^2(r)\leq\frac{(N_1+1)(N_2+1)}{N_1+N_2+1}~.
\end{equation}
Thus, for a sufficiently strong squeezing $r$ and fixed mean photon numbers $N_1$ and $N_2$, positivity under partial transposition is violated. The strength of the violation can then be used as an entanglement quantifier.

\subsection{EPR state}\label{subsec:EPR}
A special squeezed thermal state is given by the two-mode squeezed vacuum which is also referred to as twin-beam state. It is  thus characterized by $N_1=N_2=0$ and defines a symmetric two-mode Gaussian state satisfying $a=b$ (see Eqs.\,\eqref{eq:squeezedThermala} and \eqref{eq:squeezedThermalb}). By comparison with Eq.\,\eqref{eq:PPT} one recognizes immediately that a squeezed vacuum state is entangled if and only if $r\neq 0$ which is directly clear from its representation in the Fock basis \cite{GaussianGiedke}
\begin{equation}\label{eq:WernerEnergy}
 |\psi_u\rangle=\sqrt{1-u^2}\sum_{n=0}^\infty u^n |n\rangle\otimes|n\rangle~,
\end{equation}
where $u=\tanh(r)$ and $|n\rangle$ denotes the $n$th Fock state. Hence, this state is the analog of a maximally entangled state for continuous variable systems as it corresponds to the Schmidt decomposition of a maximally entangled state for $r\rightarrow\infty$. The state represents the physical realization of the model used by Einstein, Podolski and Rosen in their famous Gedankenexperiment \cite{EPR} for which reason it is also called EPR state \cite{ReviewGaussian}. This becomes clear by considering the state in the position representation which is given by
\begin{eqnarray}\label{eq:WernerWave}
 \lefteqn{ \psi_u(q_1,q_2) = } \\ 
 && \frac{1}{\sqrt{\pi}} 
 \exp\left[-\frac{1}{4}\frac{1-u}{1+u}(q_1+q_2)^2
 -\frac{1}{4}\frac{1+u}{1-u}(q_1-q_2)^2\right]. \nonumber
\end{eqnarray}
One immediately recognizes that the exponent is dominated by the second term in the limit 
$r\rightarrow +\infty$, yielding a wave function with strongly correlated particle positions. In the opposite limit, i.e.
$r\rightarrow -\infty$, the wave function describes strong 
anti-correlations of the positions of the particles. The wave function of the EPR state thus approximates a delta function $\delta(q_1\mp q_2)$ for $r\rightarrow\pm\infty$. In particular, this implies that this state approaches an eigenstate of the operator 
\begin{equation}\label{eq:Apm}
 A_\mp=\hat{q}_1\mp\hat{q}_2
\end{equation}
in the appropriate limit leading to a vanishing variance, i.e. $\langle A_\mp^2\rangle-\langle A_\mp\rangle^2\rightarrow 0$ for $r\rightarrow\pm\infty$.

\section{Probing the squeezing via backflow of information}\label{sec:backflow}
In this section we give a brief survey of the notion of non-Markovianity in the dynamics of an open quantum system quantified by an information backflow. It is furthermore shown how the non-Markovianity, which is connected to the dynamics of the Bell states in this model, can be used to probe the squeezing of squeezed thermal states. We start by recalling that the measure quantifying the degree of non-Markovianity based on the information backflow is given by \cite{Measure,OptimalPair}
\begin{align}\label{eq:measure}
 \mathcal{N}(\Phi)\equiv\max_{\rho_1\perp\rho_2}\int_{\sigma>0}\mathrm{d}t~ \sigma(t,\rho_1,\rho_2)~,
\end{align}
where $\rho_1$, $\rho_2$ are two orthogonal states of the open system and
\begin{align}
 \sigma(t,\rho_1,\rho_2)&\equiv\frac{\mathrm{d}}{\mathrm{d}t}\mathcal{D}\bigl(\Phi_t(\rho_1),\Phi_t(\rho_2)\bigr)
\end{align}
describes the rate of change of the trace distance $\mathcal{D}\bigl(\Phi_t(\rho_1),\Phi_t(\rho_2)\bigr)=\tfrac{1}{2}\mathrm{tr}|\Phi_t(\rho_1)-\Phi_t(\rho_2)|$ of these states. This can be understood as an information flux due to the interpretation of the trace distance as a measure for the distinguishability of two states \cite{Fuchs,Hayashi,Nielsen}. The set $\Phi=\{\Phi_t|0\leq t\leq T\}$ denotes the one-parameter family of dynamical maps describing the dynamical evolution of the open system. Hence, the measure $\mathcal{N}$ determines the maximal increase of distinguishability for all pairs of (orthogonal) input states.

Employing this tool to quantify memory effects in our pure dephasing dynamics \eqref{eq:genstate}, a backflow of information is signified by an increase of the coherences of the combined two-level systems. Of particular importance are the coherences $\kappa_{12}(t)$ and $\Lambda_{12}(t)$ (see Eqs.\,\eqref{eq:kappa12} and \eqref{eq:lambda12}) as they describe the nonlocal features of the joined state of the open system. It can be shown that they are connected to the entanglement of the Bell states
\begin{align}
|\Psi_I^\pm\rangle&=\frac{1}{\sqrt{2}}(|00\rangle\pm|11\rangle)~,\label{eq:psiI}\\
|\Psi_{II}^\pm\rangle&=\frac{1}{\sqrt{2}}(|01\rangle\pm|10\rangle)~\label{eq:psiII}
\end{align}
measured by the concurrence. For a two-qubit state having a spectral decomposition in terms of maximally entangled states with corresponding eigenvalues $p_i$ ($i=1,\dots,4$), the concurrence is given by \cite{Hayashi} 
\begin{equation}
 C(\rho)=\max_{i\in\{1,\dots,4\}}2p_i-1~,
\end{equation}
The entanglement of the states $\rho_{I/II}^\pm=|\Psi_{I/II}^\pm\rangle\langle\Psi_{I/II}^\pm|$ at time $t$ thus obeys
\begin{equation}
 C(\rho_I^\pm(t))=|\kappa_{12}(t)|~,\qquad C(\rho_{II}^\pm(t))=|\Lambda_{12}(t)|~.
\end{equation}
In addition, one finds for the trace distance of the orthogonal Bell states
\begin{align}
\mathcal{D}\bigl(\rho_I^+(t),\rho_I^-(t)\bigr)&=|\kappa_{12}(t)|~,\label{eq:TracePsiI}\\
\mathcal{D}\bigl(\rho_{II}^+(t),\rho_{II}^-(t)\bigr)&=|\Lambda_{12}(t)|~,\label{eq:TracePsiII}
\end{align}
so that an increase of the distinguishability of the orthogonal Bell states is linked to revivals of entanglement quantified by the concurrence in our model.
\begin{figure}[t]
    \centering
     \includegraphics[width=0.475\textwidth]{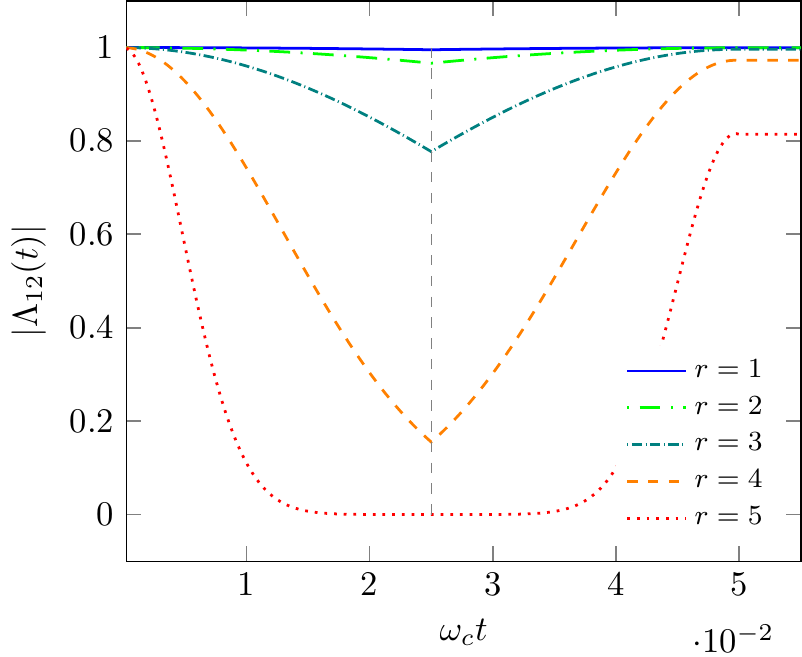}\\
    \caption{(Color online) Dynamics of $|\Lambda_{12}(t)|$ for an environmental state described by $\boldsymbol{\sigma}_{X,r}^\mathrm{EPR}$ with coupling strength $\alpha_{1,2}=1$ and subsequently applied interactions of length $2.5\cdot10^{-2}$ (in units of $\omega_c^{-1}$). Several values of the squeezing parameter $r$ for the EPR state are considered. One obtains for the measure $\mathcal{N}$: $4\cdot10^{-3}$ ($r=1$), $3\cdot10^{-2}$ ($r=2$), $0.22$ ($r=3$), $0.82$ ($r=4$), $0.82$ ($r=5$).} \label{fig:EPRGesamt}
 \end{figure}

\subsection{Nonlocal rephasing for squeezed thermal states}
We now study the dynamics of the two two-level systems given by Eq.\,\eqref{eq:genstate} induced by squeezed thermal states with a covariance matrix whose nonzero entries obey Eqs.\,\eqref{eq:squeezedThermala}-\eqref{eq:squeezedThermalc}. In the following we restrict our considerations to subsequently applied local interactions which last equally long. Performing the maximization included in the measure $\mathcal{N}$ \eqref{eq:measure} numerically, one finds that the maximal increase is given by one of the two orthogonal pairs of Bell states as the coherence factors $\kappa_1(t)$ and $\kappa_2(t)$ decay monotonically. Thus, non-Markovian behavior is exclusively related to revivals of entanglement of the Bell states being determined by the nonlocal coherence factors $\kappa_{12}(t)$ and $\Lambda_{12}(t)$. In fact, only one of these coherence factors shows revivals depending on the sign of the squeezing parameter. For a positive squeezing, the orthogonal pair of Bell states $|\Psi_{II}^\pm\rangle$ whose entanglement is quantified by $|\Lambda_{12}(t)|$ yield the maximal backflow of information. If one reverses the sign of the squeezing, the Bell states $|\Psi_{I}^\pm\rangle$ causes the maximal 
information backflow. 

One immediately infers from the structure of the nonlocal coherence factors that the nonlocal memory effects are generated by the function $f(t)$ [see Eq.~\eqref{eq:funcf}]. This function solely depends on the entries $c_\pm$ of the covariance matrix \eqref{eq:squeezedThermalc}, so that the revival of the coherences mainly depends, apart from the squeezing parameter, on the sum $N_\Sigma=N_1+N_2$ of the mean photon numbers. However, as the local factors $\kappa_1(t)$ and $\kappa_2(t)$ also enter the definition of the nonlocal coherence factors, they are also affected by the quantities $a$ and $b$ of the covariance matrix resulting in unique open system dynamics pertaining to the chosen parameters. Due to the parity of $\cosh$ and $\sinh$, changing the sign of the squeezing parameter amounts to replacing $f(t)$ by $1/f(t)$, thus, transforming $|\kappa_{12}(t)|$ into $|\Lambda_{12}(t)|$, and vice versa. This explains the observed behavior regarding the pair of states yielding the maximal information backflow.

Figure~\ref{fig:EPRGesamt} shows the dynamics of $|\Lambda_{12}(t)|$ for the squeezed vacuum for various positive values of the squeezing parameter and a fixed duration of the interactions of $2.5\cdot10^{-2}$ (in units of $\omega_c^{-1}$). One recognizes that the bigger the squeezing the stronger is the variation of the coherence factor. For example for $r=5$, after a complete decay, the modulus of the coherence factor $\Lambda_{12}(t)$ revives substantially within the second period of interaction before attaining its constant value of about $0.82$ after the second interaction has terminated. The non-Markovianity quantified by $\mathcal{N}$ obtained for this choice of the parameters is about $0.82$\,. Moreover, for a larger squeezing and a shorter interaction length, the effect is amplified yielding a completely decaying and subsequently reviving nonlocal coherence factor.

\begin{figure}
    \centering
     \includegraphics[width=0.475\textwidth]{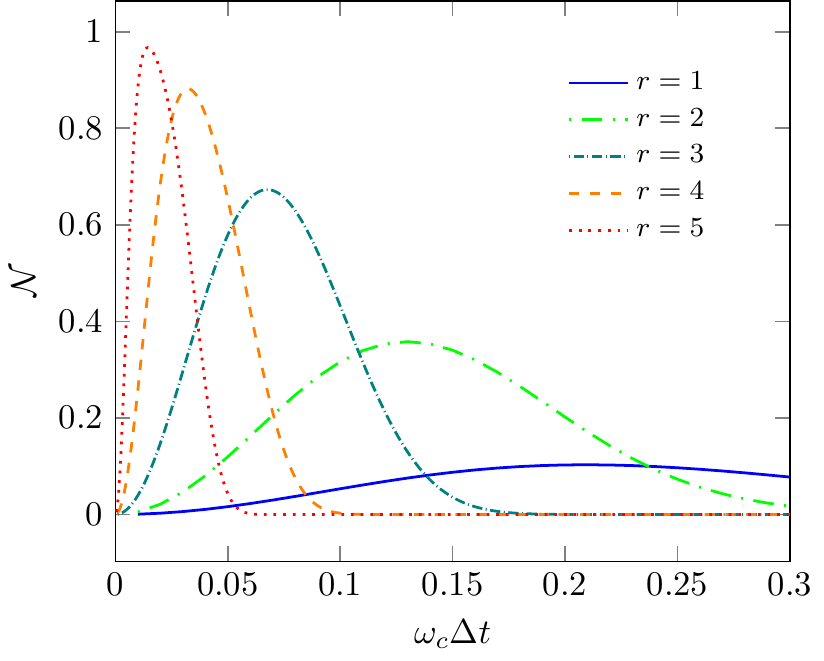}\\
    \caption{(Color online) The non-Markovianity $\mathcal{N}$ corresponding to the maximal increase of $|\Lambda_{12}(t)|$ for $\boldsymbol{\sigma}_{X,r}^\mathrm{EPR}$ as a function of the interaction length $\omega_c\Delta t$. The consecutively applied interactions are of equal length and we have chosen $\alpha_{1,2}=1$ for the coupling strength.} \label{fig:maxIncEPR}
 \end{figure}

\begin{figure}[t]
    \centering
      \includegraphics[width=0.475\textwidth]{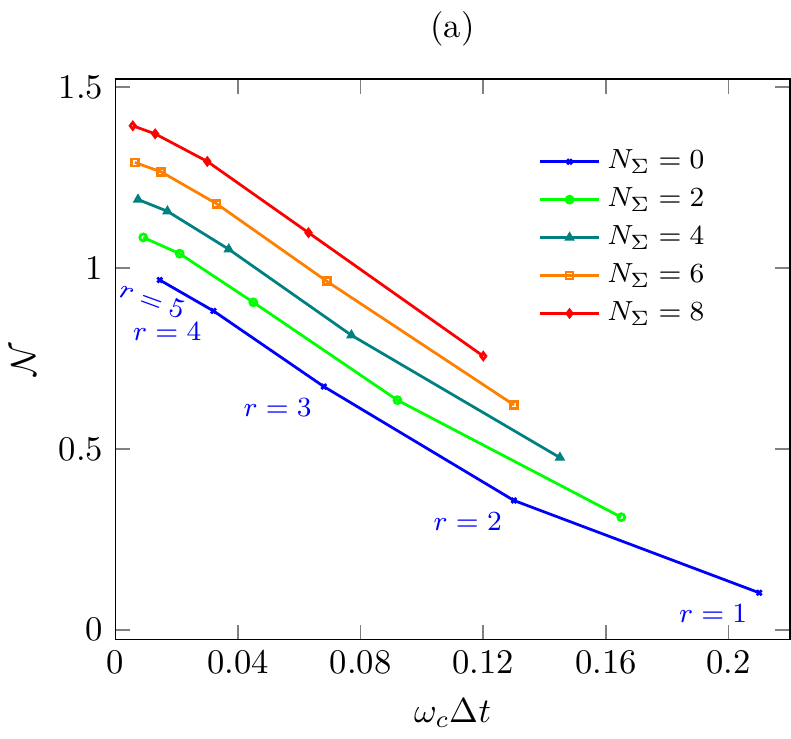}\\
      \includegraphics[width=0.475\textwidth]{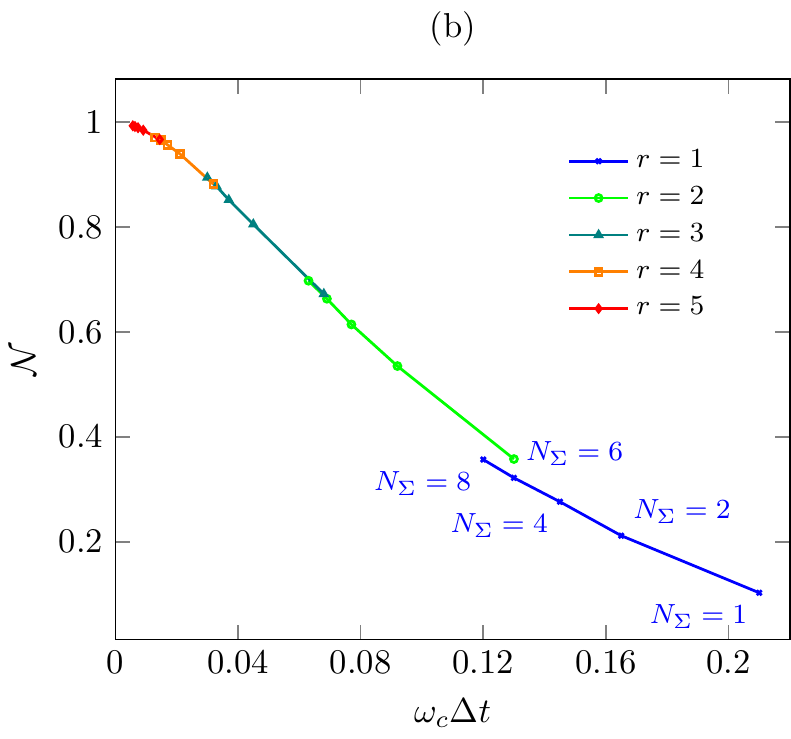}
      \caption{(Color online) The non-Markovianity $\mathcal{N}$ as a function of the optimal interaction length $\Delta t$ (in units of $\omega_c^{-1}$) inducing maximal non-Markovian dynamics for squeezed thermal states for a given value of the summed mean photon numbers $N_\Sigma$ (a) and the squeezing parameter $r$ (b). The interactions are of equal length and applied consecutively. The coupling strength is given by $\alpha_{1,2}=1$. For the sake of clarity we have chosen a stacked plot in (a) adding a cumulative offset of $0.1$ to each curve.} \label{fig:optimal}
\end{figure}

The non-Markovianity given by the maximal increase of $|\Lambda_{12}(t)|$ as a function of the interaction length for several values of the squeezing is shown in Fig.\,\ref{fig:maxIncEPR}\,. One sees that there exists an optimal duration of the subsequent interactions with respect to maximally reviving nonlocal coherences for a given value of the squeezing parameter $r$. The same observation can be made for truly thermal states where $N_i\neq0$ for $i=1,2$. Here, we concentrate on squeezed thermal states with equal mean photon numbers, i.e. $N_1=N_2$, as the dynamics of the nonlocal coherence factors is mainly affected by the sum of the mean photon numbers. Fig.\,\ref{fig:optimal} shows the optimal interaction length inducing maximal non-Markovian behavior measured by $\mathcal{N}$ for several squeezing parameters and summed mean photon numbers. Again, the stronger the squeezing for a given mean photon number $N_\Sigma$ the shorter must be the subsequently applied interactions to obtain a substantially reviving nonlocal coherence factor $\Lambda_{12}(t)$. This is reflected in the decreasing optimal interaction length depicted in Fig.\,\ref{fig:optimal}\,(a)\,. The effect of an increasing mean photon number on the optimal duration for a given squeezing can be seen in Fig.\,\ref{fig:optimal}\,(b)\,. Thus, we see that the strength of nonlocal memory effects depends on the magnitude of the coefficients $c_\pm$ \eqref{eq:squeezedThermalc} and that the optimal interaction length revealing the maximal non-Markovian behavior is a decreasing function of the magnitude of these quantities.

Furthermore, by looking at Fig.~\ref{fig:optimal}\,(a), one recognizes that the maximal non-Markovianity occurring for the optimally lasting local interactions is uniquely connected to the squeezing parameter for any fixed summed mean photon number. 
Thus, by means of the non-Markovianity of the two locally interacting two-level systems one can determine the squeezing parameter for a fixed and known temperature of the local environments. In this way the two-level systems represent a non-Markovian dynamical quantum probe for this environmental property.

An experimental implementation of this novel probing scheme requires knowledge about the mean photon numbers and to have full control over two two-level systems. This means that one must be able to prepare the system in the maximally entangled Bell states which then have to be made interacting locally with the multimode field for different times. For any duration of the interactions one must only compare the outcomes of the state tomography after the first and the second local interaction in order to determine the non-Markovianity induced by the squeezing. Hence, the scheme requires a reliable preparation of identical environmental states characterized by the squeezing parameter and mean photon numbers. We want to point out that the probing scheme is stable against an imperfect realization of equally lasting and successive local interactions. These imperfections solely diminish the rephasing of the Bell states leaving the unambiguous connection untouched which is at the heart of this strategy. A possible application of this scheme is to determine the squeezing of photon pairs generated by parametric down-conversion by means of the dynamics of the polarization degrees of freedom.

Finally, for a highly accurate experiment, it is even possible to determine the small differences in the open system dynamics caused by different mean photon numbers $N_1$ and $N_2$ which thus permits to assign not only the squeezing parameter but also the mean photon numbers to the dynamics of the Bell state. However, the accuracy regarding the determination of the non-Markovianity must be of the order of $10^{-3}$ to resolve the different evolutions which may be challenging to achieve in experiments.

\subsection{Extension to nonzero squeezing angles}
The two-mode squeezed thermal states considered previously define a special subclass of squeezed states due to the assumption of a real-valued squeezing. In general, the squeezing operator $S_2(\xi)=\exp[\xi\hat{b}_1^\dag \hat{b}_2^\dag- \xi^*\hat{b}_1\hat{b}_2]$ can have a complex valued parameter $\xi=r\exp[i\phi]$ with $\phi\in[0,2\pi)$ and $r\in\mathbb{R}_+$. The corresponding covariance matrix obtained after applying this operation to a thermal state has nonvanishing correlations between the canonical operators of the different modes, i.e. $\langle\{\hat{q}_k,\hat{p}_{j\neq k}\}\rangle\neq0$. More precisely, one finds for the covariance matrix
\begin{align}\label{eq:SqueezAngle}
 \boldsymbol{\sigma}_{X,r,\phi}^\mathrm{STS}&\equiv\frac{1}{2}\left(\begin{matrix} a I_2&c_+ R(\phi)\\c_+R(\phi) &b I_2                                                                                                                                                                                                                                                                                                                       \end{matrix}\right)~,
\end{align}
where $R(\phi)=\left(\begin{smallmatrix}\cos\phi&\sin\phi\\\sin\phi&-\cos\phi\end{smallmatrix}\right)$. Employing the odd (even) parity of $\sin$ ($\cos$), the transformation of the covariance matrix with the symplectic form $\Omega_n$ [Eq.~\eqref{eq:symplForm}] discussed in the definition of the characteristic function [Eq.~\eqref{eq:char1}] amounts to a flip of the sign of the squeezing angle $\phi$. 

The matrix $R(\phi)$ is not diagonal describing a state with nonvanishing cross-covariances if the squeezing angle is different from $0$ or $\pi$. In order to solve the dynamics of the open system in this general situation, assuming again identical states for all two-mode Gaussian state and an ohmic spectral density, one has to consider Laplace transforms $\mathcal{L}$ of $\cos$-modulated functions. For $s\neq0$ we find
\begin{align}\label{eq:cossinLa}
\mathcal{L}\left\{\cos(x t)\tfrac{\sin(y t)}{t}\right\}(s)=\frac{1}{2}\left(\arg[z_+]+\arg[z_-]\right)~,
\end{align}   
with $\arg[\cdot]$ denoting the usual argument function with values in $(-\pi,\pi]$ and
\begin{equation}
 z_\pm=1+\frac{x^2}{s^2}\pm\frac{yx}{s^2}+i\frac{y}{s}~.
\end{equation}

Employing Eq.\,\eqref{eq:cossinLa} where $s=\omega_c^{-1}>0$, the dynamics can be solved for nonvanishing cross-covariances $c_{12}$ and $c_{21}$ of a covariance matrix $\boldsymbol{\sigma}_{X}$. While the coherence factors $\kappa_{1,2}(t)$ remain unchanged, the nonlocal coherence factors attain an additional factor
\begin{widetext}
\begin{align}
 g(t)=\exp\Biggl[4\sqrt{\alpha_1\alpha_2}\Bigl\{&c_{12}\sum_{k=0}^1\bigl(\arg[1+\omega_c^2(t_2(t)+t_2^s)^2+(-1)^k\omega_c^2(t_2(t)+t_2^s)t_1(t)+i\omega_ct_1(t)]
 \nonumber \\
&\qquad-\arg[1+\omega_c^2{t_2^s}^2+(-1)^k\omega_c^2t_2^st_1(t)+i\omega_ct_1(t)]\bigr) \nonumber \\
&+ c_{21}\cdot2\bigl(\arg[1+i\omega_ct_2^s]-\arg[1+i\omega_c(t_2(t)+t_2^s)]\bigr)\Bigr\}\Biggr]~,
\end{align}
\end{widetext}
where we have again set $t_1^s=0$. For a general squeezed thermal state the nonlocal coherence factors \eqref{eq:kappa12P} and \eqref{eq:lambda12P} are then given by
\begin{align}
\kappa_{12}(t)&=\kappa_1(t)\kappa_2(t)f(t)g(t)~,\\
\Lambda_{12}(t)&=\kappa_1(t)\kappa_2^*(t)/\left[f(t)g(t)\right]~,
\end{align}
with $c_{12}=c_{21}=\tfrac{1}{2}(1+N_1+N_2)\sinh(2r)\sin\phi$.

\begin{figure}
   \centering
      \includegraphics[width=0.475\textwidth]{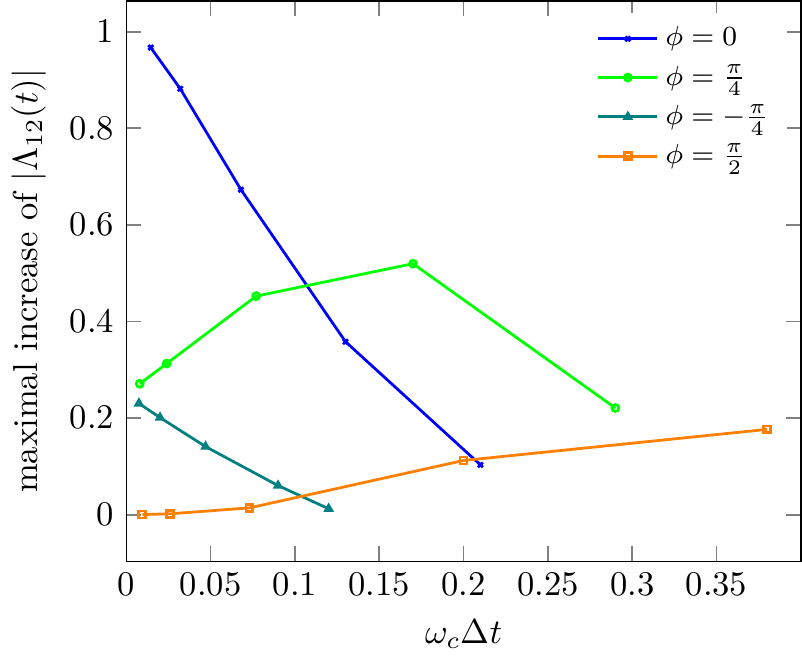}
    \caption{(Color online) The maximal increase of $|\Lambda_{12}(t)|$ as a function of the optimal interaction length $\omega_c\Delta t$ for environmental states described by $\boldsymbol{\sigma}_{X,r}^\mathrm{STS}$ with $N_1=N_2=0$ and different squeezing angles $\phi$. The data points correspond to the rephasing of $|\Lambda_{12}(t)|$ obtained for different squeezing parameters ranging from $r=1$ to $r=5$ in integer steps. The optimal interaction length again reduces with increasing squeezing parameter.} \label{fig:AngleGesamt}
 \end{figure}
 
The optimal interaction length leading to a maximal rephasing of $|\Lambda_{12}(t)|$ for nonzero squeezing angles and several squeezing parameters is depicted in Fig.~\ref{fig:AngleGesamt}. We point out that the non-Markovianity of the dynamics does not correspond to this rephasing in general. The maximizing pair might be given by the Bell states $|\psi_I^\pm\rangle$ associated to $\kappa_{12}(t)$.

Once again one observes that the larger the squeezing parameter the shorter the optimal duration of the local interactions. However, the connection between the rephasing as a function of the optimal interaction length is not as monotonic as for squeezing angles $\phi$ equal to $0$ or $\pi$\,. While the magnitude of the revivals of $\Lambda_{12}(t)$ increases continuously with the squeezing for some parameters, e.g. $\phi=-\pi/4$, it shows a non-monotonic behavior for $\phi=\pi/4$ and decays even monotonically if $\phi=\pi/2$. Nonetheless, the unambiguity of the rephasing as a function of the optimal duration of the local interactions and the squeezing parameters still allows the use of the open system as a dynamical quantum probe for these environmental properties. Hence, the monitoring of the time evolution of the Bell states for subsequently applied interactions and the determination of the resulting rephasing allows to estimate the squeezing parameter as well as the squeezing angle for any fixed and known mean photon numbers.

\section{Separable Gaussian states with strong nonlocal memory effects}\label{sec:symState}
In the preceding section we discussed the existence of memory effects associated to the rephasing of the nonlocal coherences and showed how they can be employed as a quantum probe for the squeezing of two-mode squeezed thermal states. By looking at the necessary and sufficient criterion for entanglement in these states [see Eq.\,\eqref{eq:PPT}], one observes that the considered thermal states are entangled except for the parameter combination $r=1$ and $N_\Sigma=8$. The induced non-Markovianity for this environmental state is, however, just about $0.4$ for an already relatively short optimal interaction length $\omega_c\Delta t=0.12$\,. 

Here, we show that entanglement is not necessary for strong nonlocal memory effects by considering the dynamics generated by two-mode mixed thermal states (MTS). These zero-mean Gaussian states are generated applying the unitary operator $U(\chi)=\exp[\chi \hat{b}_1^\dag\hat{b}_2-\chi^*\hat{b}_1\hat{b}_2^\dag]$ with $\chi=\tau\exp[i\theta]$, where $\theta\in[0,2\pi)$ and $\tau\in\mathbb{R}_+$, to a two-mode thermal state with mean photon number $N_1$ and $N_2$, respectively. The unitary operation is, for example, represented by the action of a beam splitter on two modes of the electromagnetic field. In this case, the quantity $\cos^2\tau$ is typically referred to as the transmissivity of the beam splitter.

Assuming a perfect beam splitter, i.e. $\cos^2\tau=1/2$, and $\theta=0$ the diagonal covariance matrix of a two-mode thermal state given by $\boldsymbol{\sigma}_{X,\mathrm{th}}=\mathrm{diag}(N_1+\tfrac{1}{2},N_1+\tfrac{1}{2},N_2+\tfrac{1}{2},N_2+\tfrac{1}{2})$ transforms according to
\begin{equation}\label{eq:SymCov}
 \boldsymbol{\sigma}_{X,N_i}^\mathrm{MTS}\equiv\frac{1}{2}\left(\begin{matrix} (N_1+N_2+1)I_2&(N_2-N_1)I_2\\(N_2-N_1)I_2&(N_1+N_2+1)I_2                                                                                                                                                                                                                                                                                                                       \end{matrix}\right).
\end{equation}
Thus, the covariance matrix is in standard form and these states are obviously mixed for $N_1,N_2\neq0$ which can also be seen by evaluating the purity, which is given by \cite{Olivares}
\begin{equation}
 \mathrm{Tr}(\rho^2)=(2^n\sqrt{\det\sigma_X})^{-1}
\end{equation}
for an $n$-mode Gaussian state. If we take the first mode entering the beam splitter to be in the vacuum state, i.e. $N_1=0$, and choose the parametrization $N_2=\cosh(2r)-1$ with $r\in\mathbb{R}$ for the mean photon number of the second mode, the covariance matrix obeys
\begin{equation}\label{eq:SymCov2}
 \boldsymbol{\sigma}_{X,r}^\mathrm{MTS}\equiv\frac{1}{2}\left(\begin{matrix} \cosh(2r)I_2&(\cosh(2r)-1)I_2\\(\cosh(2r)-1)I_2&\cosh(2r)I_2                                                                                                                                                                                                                                                                                                                       \end{matrix}\right).
\end{equation}
The passive transformation induced by $U(\chi)$ does not entangle the zero-mean two-mode thermal Gaussian state. For symmetric covariance matrices $\boldsymbol{\sigma}_{\mathrm{sym}}$ in standard form, i.e. $a=b$ in \eqref{eq:standardform1}, the Robertson-Schr\"odinger uncertainty relation \eqref{eq:conCova} and the PPT criterion \cite{GaussianZoller,GaussianSimon} can be rewritten in the following way \cite{Olivares}:
\begin{align}
 \boldsymbol{\sigma}_{\mathrm{sym}}~~ \text{physically valid}~~&\Leftrightarrow~ (a-c_+)(a-c_-)\geq\frac{1}{4},\\
\boldsymbol{\sigma}_{\mathrm{sym}}~~ \text{entangled}~~&\Leftrightarrow~ (a-c_+)(a+c_-)<\frac{1}{4},
\end{align}
whereby one deduces that the covariance matrix $\boldsymbol{\sigma}_{X,r}^\mathrm{MTS}$ defines a separable state for all $r\in\mathbb{R}$. Hence, for 
$r\neq 0$, implying nonvanishing cross-covariances $\langle\hat{q}_1\hat{q}_2\rangle$, these states represent separable but correlated two-mode Gaussian states. 

The position representation of these state is determined by means of the Weyl transform which represents the inverse of the Wigner transform. One obtains for the density matrix $\rho_{\boldsymbol{\sigma}_{X,r}^\mathrm{MTS}}(\vec{q},\vec{q}\,')=\langle \vec{q}|\rho_{\boldsymbol{\sigma}_{X,r}^\mathrm{MTS}}|\vec{q}\,'\rangle$ with $\vec{q},\vec{q}\,'\in\mathbb{R}^2$
\begin{align}
&\rho_{\boldsymbol{\sigma}_{X,r}^\mathrm{S}}(\vec{q},\vec{q}\,')\nonumber\\
&=\frac{1}{\pi}\sqrt{\frac{1}{2\cosh(2r)-1}}\exp\Bigl[-\frac{1}{4}(\vec{q}-\vec{q}\,')^T B(\vec{q}-\vec{q}\,')\Bigr]\nonumber\\
&\qquad\cdot\exp\Bigl[-\frac{1}{4}(\vec{q}+\vec{q}\,')^T B^{-1}(\vec{q}+\vec{q}\,')\Bigr]~,
\end{align}
where
\begin{equation}
B=\left(\begin{matrix}\cosh(2r)&\cosh(2r)-1\\\cosh(2r)-1&\cosh(2r)\end{matrix}\right) 
\end{equation}
For $r\gg1$ one has $B^{-1}\approx\tfrac{1}{2}\left(\begin{smallmatrix}1&-1\\-1&1\end{smallmatrix}\right)$ so that the diagonal elements $\rho_{\boldsymbol{\sigma}_{X,r}^\mathrm{MTS}}(\vec{q},\vec{q})$ of the density matrix are proportional to
\begin{equation}
 \rho_{\boldsymbol{\sigma}_{X,r}^\mathrm{MTS}}(\vec{q},\vec{q})\simeq\exp\left[-\frac{1}{2}(q_1-q_2)^2\right]~,
\end{equation}
representing a Gaussian distribution with constant variance with respect to difference in the particle positions $q_1$ and $q_2$. This is in contrast to the behavior for the EPR state where the diagonal elements in the position representation are increasingly (anti-)correlated in the particle positions with variance $\exp[-2|r|]$ for positive (negative) squeezing parameters, approaching a delta function $\delta(x_1\mp x_2)$ for $r\rightarrow\pm\infty$. The (zero-mean) two-mode mixed thermal states thus have a nonvanishing second moment of the operator $ A_-=\hat{q}_1-\hat{q}_2$ (cf. Eq.\,\eqref{eq:Apm}) for any value of $r$. More precisely, one finds $\langle A_-^2\rangle=1$.

\begin{figure}[t]
   \centering
    \includegraphics[width=0.475\textwidth]{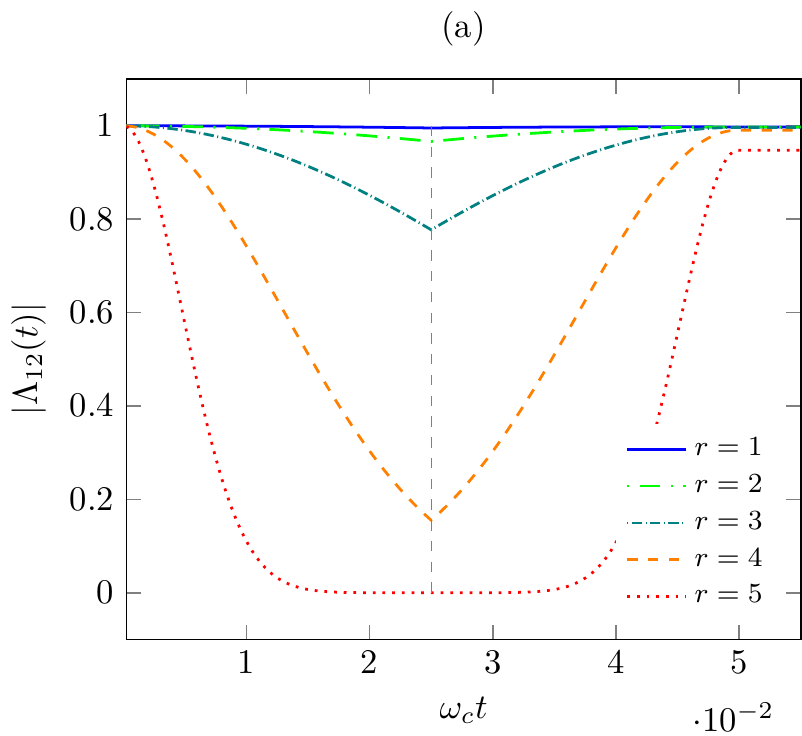}\\
    \includegraphics[width=0.475\textwidth]{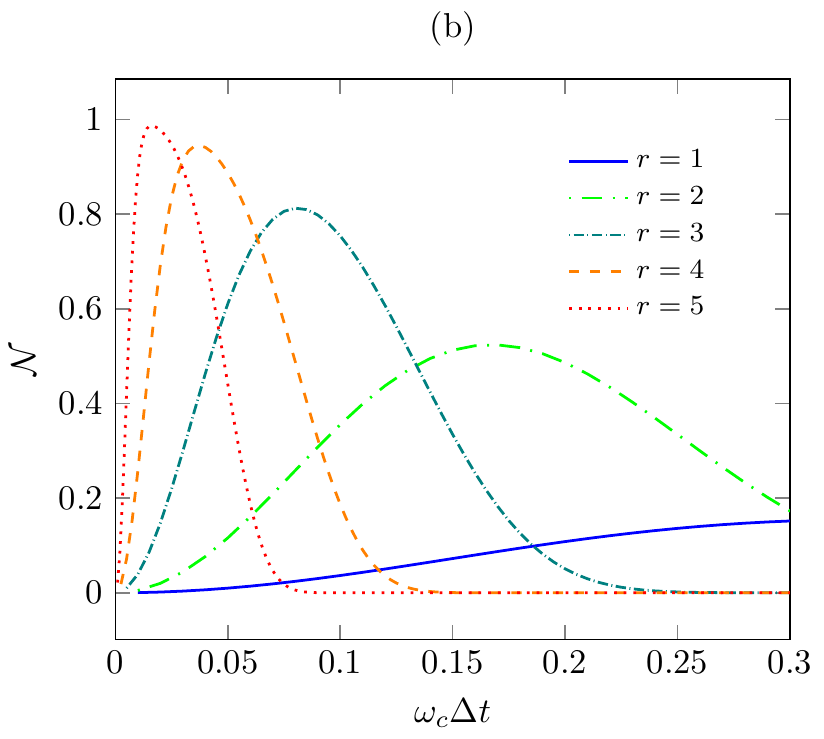}
    \caption{(Color online) (a) The same as Fig.\,\ref{fig:EPRGesamt} for environmental states described by $\boldsymbol{\sigma}_{X,r}^\mathrm{S}$. The non-Markovianity measured by $\mathcal{N}$ is given by: $4\cdot10^{-3}$ ($r=1$), $3\cdot10^{-2}$ ($r=2$), $0.22$ ($r=3$), $0.84$ ($r=4$), $0.95$ ($r=5$). (b) The optimal interaction length in terms of a maximal non-Markovian evolution for $\boldsymbol{\sigma}_{X,r}^\mathrm{S}$ as shown in Fig.\,\ref{fig:maxIncEPR} for $\boldsymbol{\sigma}_{X,r}^\mathrm{EPR}$.} \label{fig:SymGesamt}
 \end{figure}

The dynamics of $|\Lambda_{12}(t)|$ induced by these Gaussian states for $\alpha_{1,2}=1$ and subsequently applied interactions of equal length ($\omega_c\Delta t=2.5\cdot10^{-2}$) is presented in Fig.\,\ref{fig:SymGesamt}\,(a). By comparing the curves to the time evolution of the coherence factor induced by the EPR state (see Fig.\,\ref{fig:EPRGesamt}) one recognizes that the rephasing is larger for the separable Gaussian states. This observation is confirmed evaluating the non-Markovianity as a function of the interaction length depicted in Fig.\,\ref{fig:SymGesamt}\,(b). One recognizes that the optimal duration of the local interaction is longer compared to EPR states along with an amplification of the induced non-Markovianity.

Note that the maximizing pair, exhibiting the largest backflow of information, is given by the Bell states $|\Psi_{II}^\pm\rangle$ for all values of $r$ due to the symmetry of $\boldsymbol{\sigma}_{X,r}^\mathrm{MTS}$. Thus, while the entanglement of $|\Psi_{I}^\pm\rangle$ is monotonically decreasing due to the interaction it can be reobtained and thus preserved for an appropriately chosen interaction length for the other Bell states.

Though the dynamics of $|\Lambda_{12}(t)|$ induced by these separable Gaussian states is independent of the instant of time $t_1^s$ at which the first interaction is turned on, the rephasing of the coherence factor is suppressed for $t_1^s>0$ if the initial two-mode Gaussian states are chosen to be the EPR states. To explain this we observe that Gaussian states remain Gaussian under the free evolution of the bath modes, solely changing the argument of the characteristic function \eqref{eq:chiGen} according to
\begin{equation}
 \vec{z}~\mapsto~\vec{\tilde{z}}=(\exp[i\omega_1t]z_1,\dots,\exp[i\omega_nt]z_n)^T~,
\end{equation}
where $\omega_i$ refers to the frequencies of the involved mode in the Weyl operator \eqref{eq:Weyl}. For the separable Gaussian states \eqref{eq:SymCov2} one easily shows that the phase factors cancel out so that the characteristic function is invariant under the free evolution. This does, however, not hold for the EPR states. In this case the characteristic function depends on the frequencies and the elapsed time $t$ making it impossible to retrieve the same Gaussian state for all pairs of modes after the free evolution. Hence, having identical Gaussian states seems to be of great importance for the occurrence of nonlocal memory effects which is spoiled by the free evolution of the bath modes. We elaborate on this observation in the next section.

\section{Explanation of the nonlocal rephasing}

After having shown that there exist nonlocal memory effects with respect to the revivals of nonlocal coherences for appropriately adjusted parameters we now develop an explanation for the occurrence of this feature and derive a necessary and sufficient condition based on the entries of the covariance matrix.

We have observed that a proper choice of two-mode Gaussian states along with suitably lasting local interactions are essential. While for very 
short-time interactions the interesting coherence factor remains almost constant, it decays irretrievably for long-lasting ones. Furthermore, for successive interactions, introducing a gap between the turning off and on of the two local interactions, the revivals of the coherence factors are diminished, too. The same phenomenon is observed for the EPR states if the turning on of the first interaction is nonequal to zero as reported previously. Thus, the free evolution of the bath modes seems to lead to an irrecoverable displacement of the environmental states which evolve according to the dynamics induced by orthogonal states of the open system. This conjecture is strengthened by the findings for the second model presented in Ref.\,\cite{NonlocalTheo} showing nonlocal memory effects where no free evolution is incorporated. There, the rephasing relies solely on the equal duration of the subsequent interactions being independent of the actual length and a possible gap between the interactions. Furthermore, for environmental states in this model having perfectly anticorrelated frequency distributions, i.e. a correlation coefficient $K=-1$, the dynamics of the nonlocal coherence factors featuring revivals is trivial for simultaneously active interactions. The corresponding states of the environment are eigenstates of the interaction Hamiltonian if both interactions are turned on at the same time.

Based on these observation, we claim that nonlocal memory effects in the studied model can be explained by a modified ansatz neglecting the free evolution of the bath modes. For convenience, we also erase the free evolution of the open system so that the dynamics is then solely governed by $H_{int}(t)$. For two-mode Gaussian states showing nonlocal memory effects, we then expect to observe (almost) perfectly reviving coherence factors, $\Lambda_{12}(t)$ or $\kappa_{12}(t)$, for consecutive interactions while they should be (almost) constant if the interactions coexist under the approximate dynamics.

\subsection{General dephasing model}
We start by considering a general model undergoing dephasing which also comprises the approximated dynamics introduced before. The decoherence function $F(t)$ has the general structure
\begin{equation}\label{eq:genDecoh}
 F(t)=\mathrm{tr}\{\exp[iAt]\rho_E\}~.
\end{equation}
where $A$ refers to a self-adjoint operator. We assume that this operator can be written as the difference of two local operators $A_1$ and $A_2$ which are linear in the canonical variables and may vary independently in the spirit of the local interactions present in the original model. 

If the operators $A_i$ are simultaneously nonzero and $\rho_E$ refers to a Gaussian state one easily evaluates the decoherence function by means of a cumulant expansion which terminates at second order due to the linearity of $A$ in the canonical variables:
\begin{equation}\label{eq:genDecoh2}
 F(t)=\exp\left[i\langle A\rangle t -\frac{1}{2}\langle\langle A^2\rangle\rangle t^2\right]~,
\end{equation}
where $\langle A\rangle=\mathrm{tr}\{A\rho_E\}$ and $\langle\langle A^2\rangle\rangle=\langle A^2\rangle-\langle A\rangle^2$ represents the second cumulant
(variance). This shows that $|F(t)|$ is exactly equal to $1$ for all times $t$ if and only if 
\begin{equation}
 \langle\langle A^2\rangle\rangle=0~,
\end{equation}
from which one deduces that we must have
\begin{equation}\label{eq:condConst2}
 \rho_E=\Pi_a~,
\end{equation}
where $\Pi_a$ is the projection onto some subspace of an eigenspace of $A$ with eigenvalue $a$. Note that the constraint \eqref{eq:condConst2} implies that
\begin{equation}\label{eq:condComm}
 [A,\rho_E]=0~,
\end{equation}
which is, however, only a necessary but not sufficient condition for $|F(t)|=1$ in general. 

In the case of subsequently applied operators $A_1$ and $A_2$ the decoherence factor obeys 
\begin{equation}\label{eq:genDecoh3}
 |F(t)|=\exp\left[-\frac{1}{2}\langle\langle A^2_1\rangle\rangle t^2\right]~,
\end{equation}
during the first interaction of $A_1$. Summarizing the discussion in the beginning of this section, we had that a strong rephasing effect can be observed over an appropriate time interval $[0,t]$ if and only if the reviving coherence factor is almost constant for simultaneously active interactions and decays rapidly within the first interaction period for subsequent interactions. Based on this a necessary and sufficient condition for the occurrence of strong rephasing effects in the general dephasing model is given by
\begin{equation}\label{eq:genDecohCond}
 \langle\langle A^2\rangle\rangle\ll\langle\langle A^2_1\rangle\rangle~.
\end{equation}
Assuming equal second cumulants for the local operators $A_1$ and $A_2$, i.e. $\langle\langle A^2_1\rangle\rangle=\langle\langle A^2_2\rangle\rangle$, this can be reformulated as
\begin{equation}\label{eq:genDecoh4}
 1-K_{A_1,A_2}\ll\frac{1}{2}~,
\end{equation}
where $K_{A_1A_2}=(\langle A_1A_2\rangle-\langle A_1\rangle\langle A_2\rangle)/\langle\langle A^2_1\rangle\rangle$ denotes the correlation coefficient of the correlations between $A_1$ and $A_2$. In order to satisfy Eq.\,\eqref{eq:genDecoh4} the correlation coefficient hence must be positive and close to one. 

We point out that the choice of (two-mode) Gaussian states along with environmental coupling operators which are linear in the canonical observables are essential for the condition for the occurrence of strong nonlocal memory effects in the general dephasing model. If one of these requirements is not satisfied the cumulant expansion will not truncate at second order, leading in general to a more complicated set of conditions for the rephasing.

\subsection{Approximate dynamics}
Neglecting the free evolution of both the bath modes and the open system the decoherence functions for the original model are determined. For a real-valued coupling strength $g_k^i$ the evolution of the $k$th mode can be written as [see Eq.\,\eqref{eq:Hint}]
\begin{align}
 H_{int}^k(t)=\sqrt{2}\sum_{i=1}^2 g_k^i\chi_i(t) \hat{\sigma}_z^i\otimes \hat{q}_k^i~,
\end{align}
where $\hat{q}_k^i$ refers to the canonical position operator associated to the $k$th mode of subsystem $i=1,2$. Applying this Hamiltonian to the state $|11\rangle$ we see that the motion of the environmental degrees of freedoms can be described by an effective Hamiltonian
\begin{align}\label{eq:Hint11}
 H_{int}^{k,11}(t)=\sqrt{2}g_k(\chi_1(t)\hat{q}_k^1+\chi_2(t)\hat{q}_k^2)~,
\end{align}
while one obtains
\begin{align}\label{eq:Hint10}
 H_{int}^{k,10}(t)=\sqrt{2}g_k(\chi_1(t)\hat{q}_k^1-\chi_2(t)\hat{q}_k^2)~,
\end{align}
if $H_{int}^k(t)$ is applied to the state $|10\rangle$.
Here, we assumed that the coupling constants are equal, i.e. $g_k^1=g_k^2=g_k$. Apart from the function $\chi_i(t)$ the effective Hamiltonian thus depends either on the sum or the difference of the respective positions of particle $1$ and $2$ associated to the canonical operators $\hat{q}_k^i$ for $i=1,2$\,. Similarly, one finds for $|00\rangle$ and $|01\rangle$
\begin{align}\label{eq:Hint00+01}
 H_{int}^{k,00}(t)=-H_{int}^{k,11}(t)~,~~H_{int}^{k,10}(t)=-H_{int}^{k,01}(t)~.
\end{align}
If the unitary dynamics is exclusively generated by $H_{int}(t)=\sum_k H_{int}^k(t)$, the time evolution of any environmental state induced by a system state $|ij\rangle$ is described by the operator $U_{ij}(t)=\exp\bigl[-i\sum_k \int_{0}^t \mathrm{d}s H_{int}^{k,ij}(s)\bigr]$ . Due to this, the nonlocal coherence factors are given by
\begin{align}
 \Lambda_{12}(t)&=\mathrm{Tr}\bigl\{U_{01}(t)\rho_E U_{10}(t)^\dag\bigr\}\label{eq:lambda12App}\\
&=\mathrm{Tr}\bigl\{\exp\bigl[\sqrt{2}i\sum_k 2g_k (t_1(t)\hat{q}_k^1-t_2(t)\hat{q}_k^2)\bigr]\rho_E\bigr\}\nonumber
\end{align}
and
\begin{align}
 \kappa_{12}(t)&=\mathrm{Tr}\bigl\{U_{00}(t)\rho_E U_{11}(t)^\dag\bigr\}\label{eq:kappa12App}\\
&=\mathrm{Tr}\bigl\{\exp\bigl[\sqrt{2}i\sum_k 2g_k (t_1(t)\hat{q}_k^1+t_2(t)\hat{q}_k^2)\bigr]\rho_E\bigr\}~.\nonumber
\end{align}
For simultaneously active local interactions, i.e. $t_1^f=t_2^f$, the local interaction times $t_i(t)$ are proportional to $t$ for $0\leq t\leq t_1^f$. If, on the other hand, the interactions act one after the other, assuming $t_1^f<t_2^f$ for convenience, one finds $t_1(t)=t$ and $t_2(t)=0$ for all $t\in[0,t_1^f]$.

If the environmental state $\rho_E$ is chosen to be again the tensor product of identical zero-mean two-mode Gaussian states, we can apply the results of the preceding section. By setting $A_1=2\sqrt{2}\sum_k g_k\hat{q}_k^1$ and $A_2=\pm2\sqrt{2}\sum_k g_k\hat{q}_k^2$ the nonlocal coherence factors $\Lambda_{12}(t)$ ($+$) and $\kappa_{12}(t)$ ($-$) are described by the general dephasing model [Eq.\,\eqref{eq:genDecoh}] within the time interval $[0,t_1^f]$. Since the Gaussian states are assumed to have zero mean, the correlation coefficient 
is given by $K_{A_1A_2}=\langle A_1A_2\rangle/\langle A^2_1\rangle$. Furthermore, as $\rho_{E,k}$ is a Gaussian state for all modes $k$ with an associated covariance matrix in standard form [see Eq.\,\eqref{eq:standardform1}] implying that $\langle \hat{q}_k^i\hat{q}_{k'}^j\rangle=\delta_{k,k'}\langle \hat{q}_k^i\hat{q}_{k}^j\rangle$, where $\langle \hat{q}_k^1\hat{q}_{k}^2\rangle=c_+$ and $\langle( \hat{q}_k^{1})^2\rangle=a$, one finds for the expectation values of $A_1A_2$ and $A_1^2$:
\begin{align}
 \langle A_1A_2\rangle&=\pm 8 \sum_{k,k'}g_kg_{k'}\langle \hat{q}_k^1\hat{q}_{k'}^2\rangle\nonumber\\
&=\pm 8 \sum_{k}g_k^2\langle \hat{q}_k^1\hat{q}_{k}^2\rangle\nonumber\\
&=\pm8\alpha\omega_c^2c_+~,
\end{align}
and
\begin{align}
 \langle A_1^2\rangle&= 8 \sum_{k,k'}g_kg_{k'}\langle \hat{q}_k^1\hat{q}_{k'}^1\rangle\nonumber\\
&= 8 \sum_{k}g_k^2\langle \hat{q}_k^1\hat{q}_{k}^2\rangle\nonumber\\
&=8\alpha\omega_c^2a~.
\end{align}
Here, we performed the continuum limit with an ohmic spectral density $J(\omega)=\alpha \omega \exp[-\omega/\omega_c]$ resulting in $\sum_k g_k^2~\rightarrow~\int_0^\infty \mathrm{d}\omega J(\omega)=\alpha\omega_c^2$. 

The necessary and sufficient condition given in Eq.\,\eqref{eq:genDecoh4} thus reads
\begin{equation}\label{eq:condReph}
 1\pm\frac{c_+}{a}\ll\frac{1}{2}~,
\end{equation}
which proves that we observe a strong rephasing of $\Lambda_{12}(t)$ if and only if $1-c_+/a\ll1/2$ while the approximate nonlocal coherence factor $\kappa_{12}(t)$ shows revivals if and only if $1+c_+/a\ll1/2$ is satisfied. 

We note that for symmetric Gaussian states the quotient $c_+/a$ describes the correlation coefficient $K_{\hat{q}_1,\hat{q}_2}$ between the canonical operators $\hat{q}_1$ and $\hat{q}_2$, being defined by $K_{\hat{q}_1,\hat{q}_2}\equiv\langle\hat{q}_1\hat{q}_2\rangle/\sqrt{\langle\hat{q}_1^2\rangle\langle\hat{q}_2^2\rangle}$. Hence, neglecting the free evolution, strongly correlated canonical position operators implying $K_{\hat{q}_1,\hat{q}_2}\approx1$ yield the rephasing of $\Lambda_{12}(t)$ within the second interaction period while anti-correlations of the observables $\hat{q}_1$ and $\hat{q}_2$ result in a reviving coherence factor $\kappa_{12}(t)$.
 
For the EPR state (see Sec.\,\ref{subsec:EPR}) we find for the correlation coefficient
\begin{equation}\label{eq:corrCoeffEPR}
K_{\hat{q}_1,\hat{q}_2}=\frac{c_+}{a}=\tanh(2r)~,                                                                                                                                                                                                                                                                                                                                                                                                                                                                                                     \end{equation}
which approaches $\pm1$ for $r\rightarrow\pm\infty$, thus explaining the occurrence and the transition of the reviving coherence factor if the sign of the squeezing factor is changed. For this state the nonlocal memory effects are furthermore linked to the property of being an (approximate) projection onto some subspace of the kernel of $A_+$ or $A_-$. However, as the example of two-mode mixed thermal states \eqref{eq:SymCov2} shows, this can only be a sufficient condition. For this type of symmetric Gaussian states the correlation coefficient is given by
\begin{equation}\label{eq:corrCoeffSym}
K_{\hat{q}_1,\hat{q}_2}=\frac{c_+}{a}=1-\frac{1}{\cosh(2r)}~,                                                                                                                                                                                                                                                                                                                                                                                                                                                                                                    \end{equation}
which gets close to $+1$ for $|r|\rightarrow\infty$ coinciding with our observations.

\begin{figure}[t]
    \centering
     \includegraphics[width=0.475\textwidth]{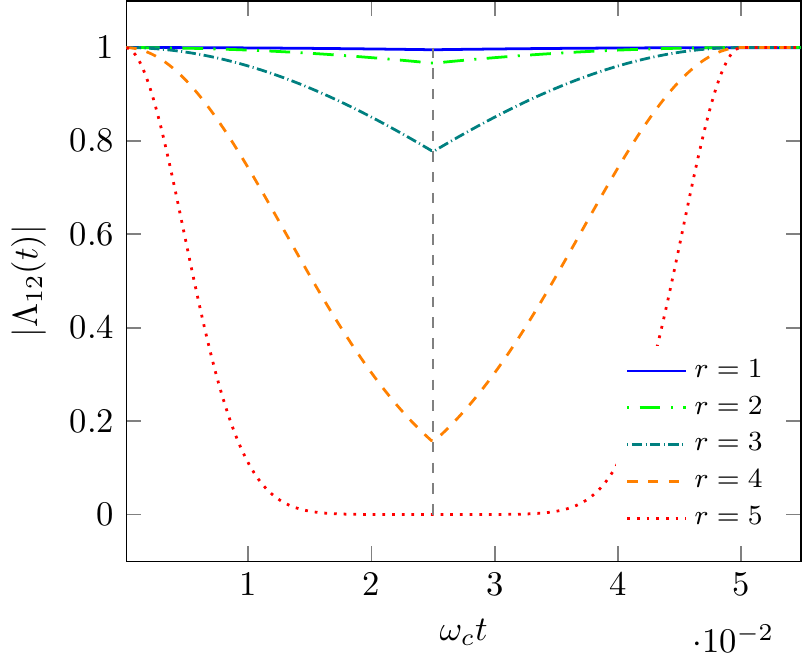}\\
    \caption{(Color online) Approximate dynamics of $|\Lambda_{12}(t)|$ for an environmental state described by $\boldsymbol{\sigma}_{X,r}^\mathrm{EPR}$ with coupling strength $\alpha=1$ and subsequently applied interactions of length $2.5\cdot10^{-2}$ (in units of $\omega_c^{-1}$). Several values of the squeezing parameter $r$ for the squeezed vacuum are considered.} \label{fig:EPRApp}
 \end{figure}
Figure \ref{fig:EPRApp} depicts the dynamics of $|\Lambda_{12}(t)|$ for successively acting local interactions and a two-mode Gaussian state described by the squeezed vacuum showing almost perfect rephasing. The coherence factors for the approximate dynamics are given by
\begin{align}\label{eq:lambda12App3}
\Lambda_{12}(t)&=e^{-4\alpha\omega_c^2 \{a(t_1(t)^2+ t_2(t)^2)-2c_+t_1(t)t_2(t)\}}~,\\
\kappa_{12}(t)&=e^{-4\alpha\omega_c^2 \{a(t_1(t)^2+t_2(t)^2)+2c_+t_1(t)t_2(t)\}}~,\label{eq:kappa12App2}
\end{align}
if an ohmic spectral density with coupling strength $\alpha$ and cutoff frequency $\omega_c$ is assumed. 

Applying the necessary and sufficient condition Eq.\,\eqref{eq:genDecohCond} to the approximate dynamics one obtains a constraint for the rephasing in terms of the first and second moments of the sum or difference of the canonical operators
\begin{equation}
 \langle (\hat{q}_1\pm\hat{q}_2)^2\rangle\ll\langle\hat{q}_1^2\rangle~,
\end{equation}
which reveals that the two classes of zero-mean two-mode Gaussian states, the squeezed vacuum and the mixed thermal states, use different mechanism leading to nonlocal memory effects. While for the mixed thermal states the expectation value of $A_-^2=(\hat{q}_1-\hat{q}_2)^2$ [cf. Eq.\,\eqref{eq:Apm}] is constant as has been mentioned in Sec.\,\ref{sec:symState}, it converges to zero for the EPR state as it defines an approximate projection onto a subspace of the kernel of $A_-$. For either states, however, the right-hand side diverges for $|r|\rightarrow\infty$.

As the diminishing effect of the free evolution is not taken into account in the derivation of Eq.\,\eqref{eq:condReph} we cannot hope that this criterion yields a sufficient condition in general if the full dynamics is considered. Despite that, we note that the maximal non-Markovianity for all considered two-mode Gaussian states is obtained for very short interaction times (cf. Figs.\,\ref{fig:maxIncEPR},\,\ref{fig:optimal} and \ref{fig:SymGesamt}(b)) so that the free evolution is almost negligible. We thus claim that under the condition $\omega_c t\ll1$ Eq.\,\eqref{eq:condReph} provides not only a necessary but also sufficient condition for the occurrence of nonlocal memory effects under the full dynamics describing the model.

The link between the occurrence of nonlocal memory effects and the correlation coefficient of the canonical operators is particularly interesting as one can show that there is a unique relation between this quantity and the generalized concurrence $C(\rho_E)$ for pure two-mode Gaussian states $\rho_E$:
\begin{equation}
 C^2(\rho_E)=1-\sqrt{1-K_{\hat{q}_1,\hat{q}_2}^2}~.
\end{equation}
Thus, entanglement amplifies the strength of the nonlocal memory effects for pure two-mode Gaussian states. However, as reported in Sec.\,\ref{sec:symState}, entanglement is not the only source for reviving nonlocal coherences. As all two-mode Gaussian states which do not factorize have nonzero quantum discord \cite{CVQuantumDiscord,CVQuantumDiscord2}, our model does unfortunately not allow to distinguish the relevance of quantum correlations for this effect.

\section{Conclusions}\label{sec:conc}
In this paper we have studied the model of two noninteracting two-level systems coupled locally to correlated bosonic environments which was introduced previously in Ref.\,\cite{NonlocalTheo} in order to analyze the emergence of non-Markovian effects induced by nonlocal environmental correlations. 

We have derived the exact solution for the dynamical map describing the time evolution of the open system for an environmental state given by a tensor product of identical two-mode Gaussian states. Our results demonstrate the occurrence of strong nonlocal memory effects due to correlated environmental states. A non-Markovian dynamics with respect to an information backflow is for example observed for squeezed thermal states. For these zero-mean two-mode Gaussian states we have shown that the strength of the nonlocal memory effects is uniquely connected to the squeezing of the multimode field as a function of the time duration of the consecutively applied local interactions. For known mean photon numbers, the dynamical evolution of the open system can thus be used to determine the squeezing in multimode fields. In this way, the locally interacting two two-level systems represent a non-Markovian dynamical quantum probe for the squeezing parameter and angle of the squeezed thermal state.

This probing scheme relies on independently switchable interactions which is motivated by the use of controllable quantum sensors testing the quantum system of interest. The non-Markovianity in the model under consideration is directly related to the evolution of the entanglement of two-qubit Bell states measured by the concurrence. The preparation of these states and the subsequent monitoring after the completion of the first and the second interaction for various interaction lengths suffices to reveal the squeezing of the two-mode Gaussian states if the mean photon numbers of the bath modes are known. We argued that the unambiguous connection of the squeezing to the revival of entanglement of a Bell state is also maintained for imperfect adjusted local interactions which overlap or are of unequal length so that this scheme can indeed be carried out at realistic experimental conditions.

Determining the dynamics caused by two-mode mixed thermal states we have furthermore demonstrated that entanglement is not necessary for the occurrence of perfectly reviving nonlocal coherences. These separable, correlated two-mode Gaussian states show even stronger memory effects than the squeezed vacuum for any time duration of the interactions. 

Finally, the phenomenon of nonlocal memory effects in this model has been explained by means of a general dephasing model and an approximate dynamics where the free evolution of the bath modes is neglected. Based on the results obtained for the general dephasing model we derived a necessary and sufficient condition for the occurrence of nonlocal memory effects for sufficiently short interaction times linking it to the magnitude of the correlation coefficient of the environmental coupling operators. We showed that the rephasing of a nonlocal coherence factor in our model occurs for short interactions compared to the environmental correlation time if and only if the correlation coefficient of the canonical position operators obeys
\begin{equation} \label{COND}
 |K_{\hat{q}_1,\hat{q}_2}|\approx 1~.
\end{equation}
The general model reveals that the assumption of Gaussian states along with environmental coupling operators which are linear in the canonical observables are essential in order to obtain this condition. 
For the EPR states condition \eqref{COND} implies that the states are
approximately given by a projection onto some subspace of the kernel of the 
interaction Hamiltonian. However, this does not hold for the separable mixed 
thermal states which are stable under the free evolution of the bath modes. In fact, 
as we have demonstrated here, these two classes of zero-mean two-mode 
Gaussian states employ quite different mechanisms leading to nonlocal memory 
effects.

\acknowledgments \label{sec:acknow}
We thank J. Piilo and E.-M. Laine for discussions. S.\,W. thanks the German National Academic Foundation for
support.

\end{document}